\RequirePackage{amsmath}
\documentclass{llncs}

\usepackage{xcolor}

\usepackage{wrapfig}
\usepackage{amsmath}
\usepackage{bm}
\usepackage{bbm}
\usepackage{amsfonts}
\usepackage[english]{babel}
\usepackage{rotating}
\usepackage{adjustbox}
\usepackage{graphicx}
\usepackage{url}
\usepackage[inline]{enumitem}

\newtheorem{observation}{Observation}

\newcommand{\mypara}[1]{\bigskip\noindent\textbf{#1.}}

\newcommand{\calR}{\mathcal{R}}

\newcommand{\prefers}[1]{\succ_{#1}}
\newcommand{\pref}{\succ}

\title{Aggregation over Metric Spaces: Proposing and Voting in Elections, Budgeting, and Legislation\thanks{A preliminary version of this work exists~\cite{preliminaryadt}.}}

\author{
  Laurent Bulteau\inst{1} \and
  Gal Shahaf\inst{1} \and
  Ehud Shapiro\inst{1} \and
  Nimrod Talmon\inst{2}
}

\institute{
LIGM, CNRS, Univ Gustave Eiffel \\
   \email{laurent.bulteau@u-pem.fr} \and
 Weizmann Institute of Science \\
    \email{$\{$gal.shahaf, ehud.shapiro$\}$@weizmann.ac.il} \and
  Ben-Gurion University \\
    \email{talmonn@bgu.ac.il}
    }

\begin{document}

\pagestyle{plain}

\maketitle

\begin{abstract}
We present a unifying framework encompassing many social choice settings. Viewing each social choice setting as voting in a suitable metric space, we consider a general model of social choice over metric spaces, in which---similarly to the spatial model of elections---each voter specifies an ideal element of the metric space. The ideal element functions as a vote, where each voter prefers elements that are closer to her ideal element.  But it also functions as a proposal, thus making all participants equal not only as voters but also as proposers. 
We consider Condorcet aggregation and a continuum of solution concepts, ranging from minimizing the sum of distances to minimizing the maximum distance.
We study applications of the abstract model to various social choice settings, including single-winner elections, committee elections, participatory budgeting, and participatory legislation.
For each setting, we compare each solution concept to known voting rules and study various properties of the resulting voting rules. %
Our framework provides expressive aggregation for a broad range of social choice settings while remaining simple for voters, and may enable a unified and integrated implementation for all these settings, as well as unified extensions such as sybil-resiliency,  proxy voting, and deliberative decision making.
\end{abstract}

\section{Introduction}

A thriving e-democracy~\cite{cacm} will have to address the choice of officers, committees, parameters (e.g., interest rate), budgets, and legislation.  Today, social choice theory addresses each of these settings separately, offering different elicitation and aggregation methods for each, thus making the practical realization of an e-democracy untenable. In addition, current theory focuses on the act of voting,  practically ignoring the need for an egalitarian process for determining which alternatives to vote upon.\emph{
Here, we present a unifying framework for all these social choice settings with a uniform elicitation and aggregation method, which is egalitarian in that voters cast both proposals and votes.}

Indeed, different models of social choice are concerned with aggregation of different objects.
For example, the most basic social choice setting is \emph{single-winner elections}
(see, e.g.,~\cite{arrowbookfirst,sen1986social,duncan1958theory}), in which voter preferences over a given set of alternatives are aggregated to decide upon a single winning alternative.
More complex social choice settings include
\emph{multiwinner elections}~\cite{mwchapter}, in which voter preferences are aggregated to select a winning committee,
and \emph{participatory budgeting} (see, e.g.,~\cite{goel2015knapsack,aussieone,abpb}), in which voter preferences are aggregated to select a bundle of items respecting a given budget limit.
Further social choice settings include graph aggregation~\cite{endriss2017graph} and aggregation of combinatorial domains~\cite{combinatorialcomsoc}.

Usually, mathematical foundations and theories in the social choice literature  are tailored for a specific setting.
Here, however, motivated by the application of a holistic e-democracy, we study a general framework of social choice that can be applied to many settings at once.
Observing the plethora of social choice settings, one merit of our approach is that it incorporates many such settings uniformly, which allows an investigation of complex settings and may enable a unified and integrated implementation to many settings at once.

Another motivation for this work comes from Reality-Aware Social Choice\cite{realsoc}, which requires the status quo (Reality) to always be present as a distinguished alternative when electing among several alternatives. According to Reality-Aware Social Choice, if the majority perceives the status quo as negative and prefers a particular alternative to the status quo, then that alternative will be chosen. 
If the majority prefers the status quo over every proposed alternative, then it will be chosen. But if the majority prefers other alternatives to the status quo, but no specific alternative among those proposed has majority support, then the status quo reigns.  This situation is clearly distressing, as the will of the people to change the status quo cannot be accommodated, perhaps due to the incompetence or interests of those who select the alternatives to vote upon, or the lack of an adequate mechanism to identify alternatives that may have majority support over the status quo. This paper proposes a model that may identify such alternatives.

A key merit of our approach is being egalitarian, in the sense that participants are not just equal as voters on proposals predetermined by others, but are also equal as proposers.  This is manifest especially in our approach to budgeting and legislation, where the domains of  budgets and laws are open-ended and potentially infinite.  This is in contrast to the classical social choice setting, in which an external authority is assumed to fix a finite number of alternatives prior to the vote. Thus, our model may enable a deliberative process in which participants transition between making proposals, negotiating changes to competing proposals to achieve common ground, building coalitions in support of a particular proposal, and supporting proposals different from their original ones as a result of negotiation, deliberation, and coalition building.  The result could be a deliberative framework in which participants are equal as proposers, deliberators, coalition builders, and voters.

Furthermore, by letting voters only specify their ideal points, and relying on the metric space to infer more complex preferences, our model can incorporate very complex aggregation scenarios (such as participatory legislation) while remaining relatively simple for voters (wrt.\ the general trade-off between expressiveness and cost). On the other hand, this simplicity means that certain voter preferences cannot be expressed, such as a dependency among budget items or the inconsistency of an intermediate point between two legislative texts; we elaborate on this point in Section~\ref{section:outlook} and discuss possible remedies. Finally, we find efficient aggregation algorithms to many of the scenarios we consider.  Coping with the NP-hardness of the remaining scenarios is left for future study.

Our model, described in Section~\ref{section:formal model}, consists of elements connected via a metric. The elements of the metric space may serve in three roles:
  proposals,
  votes,
  and outcomes of the vote aggregation process.
We associate each voter with a single element of the space, referred to as the voter \emph{ideal element}, understood as her most-preferred element of the space.
Then, we assume that each voter prefers closer elements to elements which are farther away from her ideal element.
Given a metric space, we are interested in aggregating several voter ideal elements in it into an \emph{aggregated element}, which is the election winner (we discuss other options in Section~\ref{section:outlook}).
We study several solution concepts:
  Condorcet aggregation,
  in which a Condorcet-winning aggregated element is selected (if it exists);
  and a continuum of solution concepts, containing on one end minimizing the sum of distances to the ideal elements and, at the other end,  minimizing the maximum distance to the ideal elements.

We study applications of this model to various settings of social choice. For each setting we identify a metric space suitable for the application, and investigate each solution concept by comparing it to known aggregation methods and studying its axiomatic properties.
While our model can be applied to many social choice settings, here we concentrate on the following, prominent ones:
\begin{enumerate*}

\item
Plurality elections (Section~\ref{section:plurality});

\item
Social welfare functions (Section~\ref{section:swf});

\item
Single-winner elections over a $1$-dimensional Euclidean issue space (Section~\ref{section:oned});

\item
Committee elections (Section~\ref{section:vnw});

\item
Continuous participatory budgeting (Section~\ref{section:budgeting}); and

\item
2

Participatory legislation (Section~\ref{section:legislation}), in which the task is to aggregate text drafts.
\end{enumerate*}
The study of further settings is left to future work.
A key initial motivation for our work was participatory legislation, which is perhaps the most complex setting we study here. To the best of our knowledge, here we propose the first aggregation method for this important social choice setting.

\subsection{Related Work}

The spatial model of elections~\cite{SpatialModel} considers voter ideal elements and infers voter preferences using an underlying Euclidean metric. This model was later extended to Hilbert spaces \cite{gershkov2019voting}, normed spaces \cite{peters1993generalized}, and semi-inner product spaces \cite{gershkov2018monotonic}. Here, we consider general metric spaces, including discrete and not limited to complete and well-structured ones. This approach allows us to apply this model to a broad range of social choice settings, including elections, budgeting, committee elections and legislation.

Metric preferences are quite prominent in the social choice literature. Much of this research is  similar to ours in considering ideal elements of voters and alternatives. Contrary to our model, however, extant literature is usually concerned with aggregating derived rankings rather than with directly aggregating the ideal elements themselves. Specifically, the notion of \emph{distortion}~\cite{procaccia2006distortion} is prominent (e.g.,~\cite{goel2017metric,skowron2017social,anshelevich2018approximating}).
Here, in contrast, we aim to aggregate the ideal elements and argue this to be a productive approach.
Indeed, the idea of using distances in social choice is prominent; e.g., distance-based completion principles used in combinatorial voting~\cite{combinatorialhandbook} which relate to set extensions~\cite{barbera2004ranking} and are used in committee elections. We do not aim to reinvent the wheel in the applications we consider, but rather to propose a unified model and incorporate many social choice settings at once; thus, we discuss such relations in the respective sections.

We do mention the general concept of distance rationalizability (DR)~\cite{dr}, in which voting rules are defined via consensus classes and distances.  Here, we do not consider consensus classes per se, but aggregation methods of ideal elements. Specifically, DR is concerned with summing voter distances to reach consensus in the ordinal model of single-winner elections.  This relates to our social choice maximization (specifically, the \emph{unanimous} consensus~\cite{dr}). But, notions such as Condorcet aggregation do not easily fit to DR and, furthermore, our focus is on a general metric-based model as it applies to a variety of social choice settings.

Finally, we mention the work of Feldman et al.~\cite{feldman2016voting}; specifically, their \emph{location model} is very similar to our metric-based model, but their study is focused on other aspects, such as on randomized mechanisms.  We also mention the work of Fain et al.~\cite{fain2017sequential} and Garg et al.~\cite{garg2017collaborative}, which also consider aggregation in continuous metric spaces. Additional works that tackle \textit{facility location} mostly focus on strategic voting over simple metric spaces as the line or the circle~\cite{procaccia2009approximate,meir2012algorithms,anshelevich2017randomized}.

\section{Formal Model}\label{section:formal model}

\subsection{Metric Spaces}

Our model consists of a metric space $(X, d)$, where $X$ is a set of elements and $d:X\times X\rightarrow \mathbb{R}$ is a metric function. Namely, $d$ is: (1) \textit{symmetric}, with $d(x, y) = d(y, x)$ for every pair $x, y \in X$, (2) \textit{non-negative}, with $d(x, y) \geq 0$ and $d(x, y) = 0$ if and only if $x=y$, and (3) satisfies the \textit{triangle inequality} $d(x, z) \leq d(x,y)+d(y,z)$ holds for all $x,y,z\in X$.

\subsection{Voters, Ideal Elements, and Inferred Weak Orders}

We assume $n$ voters, where voter $i$ corresponds to an element $v_i \in X$.
We refer to $v_i$ as the \emph{ideal element} of voter $i$, interpreted as her most-preferred element of the metric space. 
Based on the ideal element $v_i$, and the metric itself, we infer a ranking over the whole metric space,
where, for each two elements $x, y \in X$, voter $i$ \emph{prefers} $x$ to $y$ (i.e., ranks $x$ higher than $y$) if and only if $d(v_i, x) < d(v_i, y)$.
A voter is indifferent to elements $x$ and $y$ for which $d(v_i, x) = d(v_i, y)$.
Slightly overloading the notation, we use $v_i$ to denote both the ideal element and this inferred weak ranking of voter $i$  (the two cases will always be clear from the context).

\subsection{Aggregation Methods}

An \emph{aggregation method} is a function $\calR:X^n\rightarrow X$ that, given a metric space $(X, d)$ and $n$ ideal points $V = \{v_1, \ldots, v_n\}$ of the voters, returns an aggregated point, denoted by $\calR(V)$, which is the winner according to $\calR$ of the election corresponding to $V$ over the metric space $(X, d)$.\footnote{Indeed, due to ties, there might be co-winners. We usually do not consider these, but assume a standard arbitrary tie-breaking mechanism.}

\mypara{Condorcet Aggregation}
We adapt the Condorcet principle to our model.

\begin{definition}[Condorcet winner]
Let $(X, d)$ be a metric space and consider $n$ voters with their ideal elements and corresponding weak rankings over $X$, denoted by $V = \{v_1, \ldots, v_n\}$.
An element $x \in X$ is a \emph{Condorcet winner} wrt.\ $V$ if
for any other element $y \in X$, it holds that $|\{ v_i \in V : x \prefers{i} y \}| > |\{ v_i \in V : y \prefers{i} x \}|$.
\end{definition}

That is,
a Condorcet-winning element is such that, if chosen, then there is no voter majority for changing it. For simplicity, we shall use the notation $V(x \prefers{} y):= |\{ v_i \in V : x \prefers{i} y \}| > |\{ v_i \in V : y \prefers{i} x \}|$, under which $x\in X$ is a Condorcet winner if $V(x \prefers{} y) > V(y \prefers{} x)$ for every $y \in X$. 
An aggregation method $\calR$ is \emph{Condorcet-consistent} if it elects a Condorcet winner whenever such exists; that is, $\calR(V) = c$ where $c$ is a Condorcet winner.\footnote{Weak Condorcet winners are defined similarly, by replacing $>$ with $\geq$ in the definition above.
As the distinction between Strong/Weak Condorcet winners is in essence an issue of tie breaking, we will not focus on it.}

\mypara{$\bm{L_p}$ Aggregation}
Besides Condorcet aggregation, we consider \textit{$L_p$ aggregation}, a continuum of solution concepts, parameterized by a parameter $1\leq p \leq \infty$. Specifically, for a given $p$, we look for an element which minimizes the sum of $p$-factored distances to the voter ideal elements.
\begin{definition}
Let $(X, d)$ be a metric space and consider $n$ votes with their ideal elements, denoted by $V = \{v_1, \ldots, v_n\}$. The $L_p$ estimator of $V$ is defined by 

$$L_p(V):=argmin_{x\in X} \sum_{i \in [n]} d(v_i, x)^p.$$
Similarly, we define $$L_\infty(V):=argmin_{x\in X} \max_i \{d(x,v_i)\}.$$
The \emph{$L_p$} aggregation method returns these elements $x$ as the co-winners.
\end{definition}

Note that $L_p(V)$ is not necessarily unique, thus we treat this term as a subset of $X$. (Furthermore, $L_p(V)$ might be empty.) To tackle the issue of uniqueness, we sometimes consider \textit{reduced $L_p$ aggregation}:

$$\widetilde{L_p}(V): = \lim_{\substack{p\neq q \\ q\rightarrow p}} L_q(V)$$

which satisfies $\widetilde{L_p}(V) \subseteq L_p(V)$ due to the continuity of $\sum_{i \in [n]} d(v_i, x)^p$.\footnote{Note that the limit in the definition of $\widetilde{L_p}(V)$ is taken over sets, i.e., $x\in \widetilde{L_p}(V)$ if there is a sequence $x_n\rightarrow x$ with $x_n\in L_{q_n}(V)$ and $q_n\rightarrow p$.} These notions would mainly be applied to resolve uniqueness issues for $p=1,\infty$.

$L_p$ norm estimators have been widely studied in the statistical literature~\cite{sposito1990some}, and are usually applied when estimating the parameters in linear regression models. The choice of the parameter $p$ in this context critically depends on the underlying distribution, and reflects a tradeoff between the number of samples needed (the \textit{efficiency} of the estimator) and its sensitivity to departures from normality in the residual distribution (the estimator's \textit{robustness}) \cite{pennecchi2006between}.

To better understand the notion of $L_p$ estimation, observe that $L_1(V)$ minimizes the \textit{sum of absolute errors}, $L_2(V)$ minimizes the \textit{mean squared error}, and $L_\infty(V)$ minimizes the maximal distance from all the ideal points.\footnote{In the statistical context there is an underlying assumption of some unknown distribution; this corresponds to objective social choice, in which the goal of voting rules is to recover some assumed ground truth~\cite{moulin2016handbook}. Here we are interested in subjective social choice, in which the goal of voting rules is to aggregate individual opinions of agents~\cite{caragiannis2017subset}; thus, we do not speak of $L_p$ estimators but of $L_p$ aggregators.}
$L_p$ norms have been also studied within the social choice literature (e.g.,~\cite{paths}).
Intuitively, in the context of vote aggregation, the choice of the parameter $p$ reflects the relative influence of an individual in a joint decision making. For example, applying $L_\infty$ as a voting rule, intuitively, protects minority rights, in the sense that it does not disregard outliers (in this context we mention the work of Aziz et al.~\cite{aziz2018egalitarian} which use the maximum operator to achieve the same protection for minorities).
Further intuition regarding the role of the parameter $p$ in the aggregation appears in Section~\ref{section:oned}.

\subsection{Axiomatic Properties}

In subsequent sections we consider applications of our general model. To better understand how different solution concepts behave in each application, we consider several axiomatic properties. 

\begin{definition}
An aggregation method $\calR$ is \emph{majoritarian} if, whenever a voter majority selects the same ideal element $w$, then $w$ is selected by the aggregation method; formally, and as there could be co-winners, $\calR$ is majoritarian if $|\{v_i \in V : v_i = w\}| \geq |V| / 2$ implies that $w \in \calR(V)$.
\end{definition}

\begin{definition}
An aggregation method $\calR$ is \emph{monotone} if, for each set $V = \{v_1, \ldots, v_{n - 1}, v\}$ of votes and for each co-winner $w \in \calR(V)$ it holds that $w \in \calR(V')$, where $V' = \{v_1, \ldots, v_{n - 1}, v'\}$ and $v' = w$.
\end{definition}

That is, majoritarity holds if the aggregated point equals the majority ideal point, while monotonicity holds if moving an ideal point to the aggregation point does not change the aggregation point.

\subsection{Overview of our Results}

Throughout, we study properties of Condorcet-aggregation and $L_p$ aggregation for certain applications of our general model.  We note that distance metrics may not be sufficiently expressive in general, as they cannot express dependencies (among candidates or budget items) or semantic relations (among sentences in a legislation). Overcoming this limitation is further discussed in Section~\ref{section:outlook}.
Table~\ref{table:allresults} lists our main results.

\begin{table}[t]
%\begin{sidewaystable}
\centering
\caption{Summary of our main results. Each block of rows corresponds to a different application, while each row corresponds to a different aggregation method. For each application we consider existence, uniqueness, majoritarity, and the computational complexity of each aggregation method.}
\label{table:allresults}
\adjustbox{width=\textwidth}{
\begin{tabular}{c c | c c c c c}
Setting & Aggregation & Solution & Unique & Complexity & Majoritarian & Monotonicity \\ \hline
\hline
Plurality
    & Condorcet
        & Plurality
        & no
        & linear
        & yes
        & yes
        \\
%SW
Elections
    & $\widetilde{L_p}(V)$
        & Plurality
        & no
        & linear
        & yes
        & yes
        \\
\hline
1D
    & Condorcet
        & Median
        & for $n \in \mathbb{N_{\textrm{odd}}}$
        & linear
        & yes
        & yes
        \\
%1D
Single
        & $\widetilde{L_p}(V)$ 
        & Median for $p = 1$
        & yes
        & linear for $p = 1, 2, \infty$
        & only for $p = 1$
        & only for $p = 1$
        \\
Winner
    & 
        & Average for $p = 2$
        & 
        & efficient for $p \neq 1, 2, \infty$
        & 
        & 
        \\
    & 
        & Mid-range for $p = \infty$
        & 
        & 
        & 
        & 
        \\
\hline
Continuous  
    & Condorcet
        & no
        & no
        & efficient
        & yes
        & yes
        \\
Budgeting
    & $\widetilde{L_p}(V)$ 
        & yes
        & yes
        & linear for $p = 2$
        & only for $p = 1$
        & no
        \\
    & 
        & 
        & 
        & efficient for $p \neq 2$
        & 
        & 
        \\
\hline
%SWF
Social
    & Condorcet
        & no
        & no
        & NP-hard
        & yes
        & yes
        \\
%SWF
Welfare
    & $\widetilde{L_p}(V)$
        & yes
        & no
        & NP-hard
        & only for $p = 1$
        & only for $p = 1$
        \\
Functions
   & 
        & 
        & 
        & 
        & 
        & 
        \\
\hline
Committee 
    & Condorcet
        & no
        & no
        & NP-hard
        & yes
        & yes
        \\
%VNW
Elections
    & $\widetilde{L_p}(V)$
        & yes
        & for $p = 1$, $n \in \mathbb{N_\textrm{odd}}$
        & linear for $p = 1$
        & only for $p = 1$
        & only for $p = 1$
        \\
    & 
        & 
        & 
        & NP-hard for $p > 1$
        & 
        & 
        \\
\hline
Participatory 
    & Condorcet
        & no
        & no
        & NP-hard
        & yes
        & yes
        \\
%Legislation
Legislation
    & $\widetilde{L_p}(V)$
        & yes
        & no
        & NP-hard for $p = 1, \infty$
        & only for $p = 1$
        & only for $p = 1$
\end{tabular}
}
\end{table}
%\end{sidewaystable}

\section{Plurality Elections}\label{section:plurality}

We apply our model to the standard, general model of single-winner elections~\cite{arrowbookfirst,sen1986social,duncan1958theory}, in which the aggregation method decides on a single winner.
Specifically, we consider the categorical setting, with no structure among the alternatives.
Thus, in this setting we have an underlying equidistant set of alternatives, and the task is to aggregate voter preferences to select one of these alternatives.
In a sense, this is the simplest social choice setting which we present also as a warm-up.

\mypara{Formal model}
Let $A$ be the underlying set of alternatives and notice that the set of possible outcomes of a single-winner election over $A$ is exactly $A$. Thus, we consider the metric space $(X, d)$ where $X = A$.
Here we assume no structure or dependencies between the alternatives, hence we define $d$ to be the \emph{discrete metric}, i.e., for each $x \neq y \in X$, $d(x, y) = 1$, and $d(x, x) = 0$. 

\smallskip

For this simple scenario, Condorcet and $L_p$ aggregation for any $1 \leq p \neq \infty$ essentially boil down to Plurality, while $L_\infty(V)$ equals Plurality for unanimous elections, but otherwise selects all alternatives as co-winners.

\begin{proposition}
  In the application to Plurality elections, Plurality is both Condorcet-consistent and satisfies $L_p(V)$, for any $p \geq 1$.
\end{proposition}

\begin{proof}
For each $x\in X$, let $V(x):=|\{ v_i \in V : v_i = x \}|$ denote the number of votes for $x$ among the votes $V=\{v_1, \ldots, v_n\}$. Let  $z\in X$ be a Plurality winner, i.e., every $x\in X$ satisfies $V(x)\leq V(z)$. 
Then, for any $x \in X$, it follows that:
\begin{enumerate}
    
    \item $V(x \prefers{} z) = V(x) \leq V(z) = V(z \prefers{} x)$, 
    
    \item $\sum_{i \in [n]} d(v_i, x)^p = n - V(x)\geq n-V(z) = \sum_{i \in [n]} d(v_i, z)^p$.
    
\end{enumerate}

Item 1. above completes the proof for Condorcet aggregation, while Item 2. above completes it for $L_p$ aggregation for $1 \leq p \neq \infty$
%.
For $p=\infty$,
notice that $z\in L_\infty(V)$ for unanimous elections (in which all voters vote the same); however, as soon as there are at least two different ideal points, we have $\max_i \{d(x,v_i)\} = 1$ for every $x\in X$, hence $L_\infty(V) = X$, thus all alternatives tie as co-winners.
\qed\end{proof}

We note that, while $L_\infty (V)$ does not contain much information here, its reduced form $\widetilde{L_\infty}(V) := \lim_{p\rightarrow \infty} L_p(V)$ is restricted to the plurality elements in~$X$.
The following corollary is immediate.

\begin{corollary}
  In the application to Plurality elections, both Condorcet aggregation and $L_p(V)$ exist for any $p \geq 1$, but are not always unique.
  Furthermore, they are monotonic, majoritarian, and may be computed in linear time.
\end{corollary}

\begin{proof}
Uniqueness is not guaranteed in situations where more than one alternative gets the highest number of votes.
Monotonicity holds as a Plurality winner remains so whenever ballots are changed to it.
Majoritarity holds as a majority of ideal elements on one alternative would make it the Plurality winner.
Plurality can be computed in linear time by summing the tally of each alternative.
\qed\end{proof}

\section{$1$-Dimensional Single-Winner Elections}\label{section:oned}

Next we consider a different model of single-winner elections, corresponding to choosing a parameter in a $1$-dimensional space (say, a price for a commodity, a tax rate) or political elections over a single political axis. We assume that some subset $X\subseteq \mathbb{R}$ is given as the set of alternatives. As is the case with Plurality elections (Section~\ref{section:plurality}), the task here is to select a single alternative as a winner.

\mypara{Formal model}
We consider a metric space $(X, d)$, where $X \subseteq \mathbb{R}$ is a set of real numbers, and $d$ corresponds to the absolute distance; i.e., for each $x, y \in X$, $d(x, y) = |x - y|$ (recall that each $x \in X$ corresponds to a real number). For the ease of notation, we shall order the ideal elements in $V$ by their magnitude, i.e., $v_1\leq v_2 \leq ...\leq v_n$.
We begin with a simple characterization of Condorcet winners and $L_p$ aggregation for $p=1,2,\infty$ in this setting.

\begin{observation}
  For the application to $1$-dimensional single-winner elections, the Condorcet aggregation and the $L_1$ aggregation are the median of $V$.\footnote{If there are two medians, then any point between them, including them, is a (weak) Condorcet winner and also minimizes $L_1$.} Furthermore, if $A\subseteq \mathbb{R}$ is convex, then $L_2(V) = \frac{1}{n}\sum_i v_i$ (the average of $V$), and $L_\infty(V)=\frac{v_1+v_n}{2}$ (its mid-range).
\end{observation}

\begin{proof}
For Condorcet aggregation, the result follows Black's celebrated Median Theorem~\cite{blackmedian}. The results for $L_p(V)$ are folklore and appear, e.g., in~\cite{pennecchi2006between}.
%
% the result is folklore; for completeness, notice the following:
%
% The function $f(x):=\sum_i|x-v_i|$ is continuous and piecewise linear, hence differentiable in $\mathbb{R}\setminus \{v_1,...,v_n\}$. Its left-most slope is $-n$. By induction, the slope increases by $2$ for each interval from left to right, with right-most slope $+n$. Hence the piece-wise slope first reaches either $−1$ or $0$ at $v_{\left \lfloor{n+1/2}\right \rfloor}$ and $0$ or $+1$ at $v_{\left \lceil{n+1/2}\right \rceil}$. Hence the function attains its minima in the interval $[v_{\left \lfloor{n+1/2}\right \rfloor},v_{\left \lceil{n+1/2}\right \rceil}]$, which reduces to a singleton when s is odd.
%
\qed\end{proof}

When $X\subseteq \mathbb{R}$ is compact, the existence of $L_p$ aggregation for other values of $p$ follows since $\sum_{i \in [n]} |v_i - x|^p$ is continuous, and thus must obtain a minimum in $A$. Furthermore, if $X\subseteq \mathbbm{R}$ is convex and $1 < p \leq \infty$, this function is strictly convex and thus obtains a unique minimum. It follows that the aggregation point for convex domains may be efficiently computed via gradient descent methods.
Otherwise, if $X$ is not convex, then an efficient algorithm can be realized by first computing $L^{conv}_p(V)$, the aggregated element wrt. the convex hull $conv(X):=[v_1,v_n]$. The convexity of $\sum_{i \in [n]} |v_i - x|^p$ implies that $L_p(V)$ is obtained in one of both points in $X$ closest to $L^{conv}_p(V)$, i.e., either $\max \{x\in X| x \leq L^{conv}_p(V)\}$ or $\min \{x\in X| x \geq L^{conv}_p(V)\}$.
We thus conclude the following.

\begin{observation}
  For the application to $1$-dimensional single-winner elections, Condorcet aggregation and $L_p$ aggregation exist wrt. any compact set of alternatives $X\subseteq \mathbb{R}$. These aggregation rules can be efficiently computed via gradient descent schemes ($L_1$ aggregation and Condorcet aggregation can also be computed in linear time, as they are equivalent to the median). If $X\subseteq \mathbbm{R}$ is convex then $L_p(V)$ is unique for every $1 < p \leq \infty$. 
\end{observation}

\begin{figure}[t]
    \centering
    \begin{minipage}[b]{.5\textwidth}
    \includegraphics[scale=0.425]{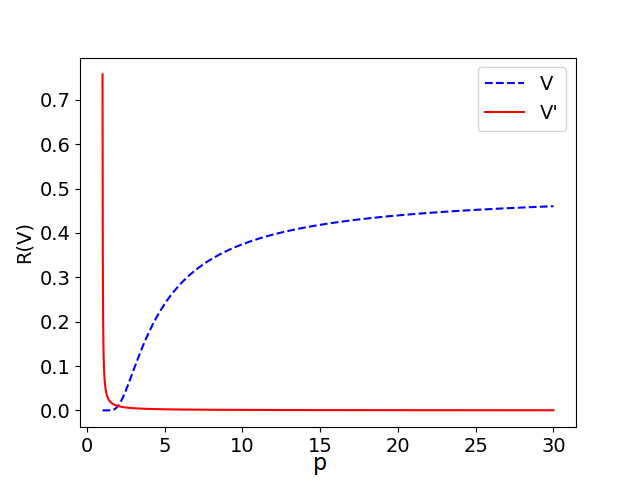}
    \end{minipage}\hfill
    \begin{minipage}[b]{.35\textwidth}
    $$V=\begin{cases}
  0 &v_1,...,v_{n-1}\\
  1           &v_n
\end{cases}$$  
$$V'=\begin{cases}
  -1 &v_1,...,v_{(n-1)/2}\\
  1           &v_{(n-1)/2}+1,...,v_{n} \\ 
\end{cases}$$ \ \\ \ \\ \ \\ 
    \end{minipage}\hfill
    \begin{minipage}[b]{.1\textwidth}
    \end{minipage}\hfill
    \caption{The figure illustrates the influence of outliers under $L_p$ aggregation as a function of $p$. Specifically, we consider two vote distributions on $X=[-1,1]$, namely $V$, $V'$ as defined above (assume $n$ is odd; we use $n = 101$). These distributions correspond to the influence of a single voter who voted 1 wrt. consensus (in $V$) and symmetric polarization ($V'$) in the rest of the votes. The blue dashed curve corresponds to $L_p(V)=\frac{1}{1+n^{1/(p-1)}}$, while the red curve corresponds to $L_p(V')=\frac{1-(\frac{n}{n+2})^{1/(p-1)}}{1+(\frac{n}{n+2})^{1/(p-1)}}$. Note that there is intersection in $p=2$ as $L_2(V)=\frac{1}{n}$ for both vote distributions.}
    \label{figure:plot}
\end{figure}

While the median, $L_1(V)$, is not necessarily unique when $|V|=n$ is even, its reduced form, $\widetilde{L_1}(V) := \lim_{p\rightarrow 1} L_p(V)$, is unique on convex domains (in a similar manner to the case of $p=\infty$  presented in Section \ref{section:plurality}).
$L_\infty(V)$ always exists, but it is not always unique.
To get some more intuition for the behavior of $L_p$ aggregation, consider the extreme cases depicted in Figure~\ref{figure:plot}.

\begin{corollary}
  For the application to $1$-dimensional single-winner elections, Condorcet aggregation and $L_p$ aggregation are not always unique. The reduced form $\widetilde{L_1}(V)$ is unique on convex domains.
\end{corollary}

Next we consider monotonicity and majoritarity.

\begin{proposition}\label{proposition:two}
  For the application to $1$-dimensional single-winner elections, Condorcet aggregation and $L_1(V)$ are majoritarian and monotonic for odd number of voters, but not monotonic for even number of voters, while $L_p$ aggregation for $p > 1$ is neither monotonic nor majoritarian.
\end{proposition}

\begin{proof}
Condorcet aggregation and $L_1(V)$ both choose the median point.
Thus, majoritarity holds as having a majority of the ideal points on one element would make it be the median.
For odd number of voters, the median is indeed a point, thus monotonicity holds as moving elements to the median would not change the median.
For even number of voters, there is not only a single median, thus monotonicity (as defined), due to tie breaking, is not satisfied.

For $L_p$ aggregation for $p > 1$, the distribution~$V$ in Figure~\ref{figure:plot} provides a counterexample to majoritarity. Monotonicity does not hold for similar reasons (e.g., moving one voter from $0$ to the aggregation point of $V$ would result in changing the aggregation point).
~\qed\end{proof}

\section{Continuous Participatory Budgeting}
\label{section:budgeting}

Here we apply our general model to a continuous model of budgeting.
Several budgeting methods exist in the literature~\cite{abpb,aussieone,donor}; we also mention related work on probabilistic social choice~\cite{probchapter}.
Here we concentrate on a model, termed \emph{continuous participatory budgeting}, in which fractional funds shall be allocated to several divisible items (in contrast to the non-divisible model of participatory budgeting~\cite{cabannes2004participatory}, in which items come with cost, and an item cannot be partially funded):\footnote{This model is sometimes also referred to as Pie Chart voting~\cite{dominikthesis}.}
  Specifically, individual financial statement items are grouped by cost centers or departments, and there is a given, fixed budget limit. In contrast to related models~\cite{garg2017collaborative,fain2016core}, we do not assume a fixed set of alternatives provided to the agents to choose from. Rather, every agent may \textit{propose} a set of alternatives, (represented, e.g., by a string) and an allocation of funds between them.

\mypara{Formal model}
We assume that each agent $i$ proposes a finite set of alternatives~$A_i$ and a distribution of the funds between these alternatives. For normalization, let the given amount be $1$, and thus the distribution corresponds to a probability vector $v_i\in \mathbb{R}^{|A_i|}$. Let $A=\bigcup_i A_i$ denote the (finite) set of alternatives proposed by the agents. We shall decide how to distribute a given amount of available funds between $|A|=m$ alternatives. 
Thus, the set of possible ideal points consists of the $(m-1)$-dimensional simplex
$$\Delta_{m-1}:=\{(v_1,...,v_m)\in \mathbb{R}^m \>|\> v_i\geq 0 \> \text{and} \> \sum_i v_i = 1\}\ ,$$
where each vector in $\Delta_{m-1}$ corresponds to a distribution of the budget among the $m$ alternatives. In order to define distances in this context, it is natural\footnote{Other options exist, such as, e.g., considering metrics induced by other norms.} to consider the metric induced by the Euclidean norm 
$$d(x,y)= \left\| x - y \right\|= \left( \sum _i |x_i-y_i|^2 \right) ^{1/2}.$$

Utilizing Plott's theorem \cite{plott1967notion}, that provides a full characterization of Condorcet winners in euclidean spaces, we conclude the following:

\begin{corollary}
  In this setting, Condorcet winners do not always exist, and their existence can determined in polynomial time.  
\end{corollary}

In fact, as follows from the literature on spatial models~\cite{SpatialModel}, a Condorcet winner exists with probability zero, and  the Condorcet cycle usually spans over almost the entire domain. 

In contrast, with regards to $L_p$ aggregators, since $f(x)=\sum_{i \in [n]} \left\| x - v_i \right\|^p$ is continuous over the compact domain $\Delta_{m-1}$, it follows that there always exists an element in $X=(\Delta_{m-1},\left\| \cdot \right\|_p)$ that minimizes $f$. Furthermore, when the ideal points $v_1,...,v_n$ are not collinear (i.e., they do not lie on the same line), it holds that $f$ is positive and strictly convex in $\Delta_{m-1}$, hence obtains a unique minimum (we refer the reader to~\cite{vardi2000multivariate} for elaborate discussion for the case of $p=1$).

\begin{corollary}
  $L_p$ aggregation always exists, and it is unique (except for the case of $p=1$ with respect to an even number of ideal points that are collinear). In contrast, Condorcet aggregation usually does not exist and is not necessarily unique.
\end{corollary}

\begin{remark}
We note that the $L_1$-estimator in this setting is the \textit{geometric median} of $v_1,...,v_n$ (sometimes called the \textit{Fermat} or the \textit{Torricelli} point), and the $L_2$-estimator is the \textit{center of mass} of the given ideal points, which corresponds to the average in every coordinate.
\end{remark}

The situation here generalizes the situation of $1$-dimensional Euclidean elections described in Section~\ref{section:oned}, as a $1$-simplex is a $1$-dimensional Euclidean space. From similar reasons as discussed in Section~\ref{section:oned} (specifically, the convexity and compactness of $\Delta_{m-1}$), efficient computation for $L_p$ aggregation follows by using gradient descent.
 
\begin{corollary}
  $L_p$ aggregation can be computed efficiently using gradient descent.
\end{corollary}

Majoritarity holds for Condorcet aggregation and for $L_1$ aggregation.

\begin{proposition}
  For the application to continuous budgeting, majoritarity holds for $L_1$ but not for $L_p$ for $p > 1$. No $p \geq 1$ satisfies monotonicity. Condorcet aggregation satisfies both monotonicity and majoritarity.
\end{proposition}

\begin{proof}
For $p > 1$ majoritarity fails due to Proposition~\ref{proposition:two}. For $p = 1$, majoritarity holds as any deviation from the majority ideal point would increase the distance from the majority voters, while, following the triangle inequality, the decrease from other voters would not suffice to account for the increase. Monotonicity is not satisfied even for $p = 1$ (e.g., consider a triangle).
\qed\end{proof}

\section{Social Welfare Functions}\label{section:swf}

We consider social welfare functions, where the input for an aggregation method is a set of $n$ linear orders and the output of an aggregation method is another linear order. A relevant metric space in this setting is the set of all permutations $X=S_z$ over a set of $z$ underlying alternatives, equipped with the metric $d$ induced by swaps of consecutive alternatives (other options are natural for future study).

\begin{observation}
    For the application to social welfare functions $L_p$ aggregation for any $1\leq p \leq \infty$ always exists but are not always unique. 
    Condorcet aggregation does not always exist.
\end{observation}

\begin{proof}
Being the minimum of a continuous function over a compact domain, the existence of $L_p$ aggregation in this setting is guaranteed. To see why they are not unique, consider a finite set of alternatives $A$ and $|A|!$ voters, each of which corresponds to a different permutation over $A$. By symmetry considerations, $L_p(V) = A$ for every  $1\leq p \leq \infty$. 

For Condorcet, consider $A=\{a,b,c\}$ with votes $v_1 : a \pref b \pref c$, $v_2 : b \pref c \pref a$, and $v_3 : c \pref a \pref b$. Then, e.g., $a \pref b \pref c \to a\pref c\pref b \to b\pref c\pref a \to b\pref a\pref c \to c\pref a\pref b \to c\pref b\pref a \to a\pref b\pref c$ is a cycle.
\end{proof}

\begin{observation}\label{observation:kemenynphard}
  In the application to social welfare functions, the $L_1$ aggregation point is the Kemeny ranking and thus $L_1$ aggregation is NP-hard.
\end{observation}

\begin{proof}\label{proposition:kemeny}
Given $n$ ordinal ballots (i.e., $n$ linear orders) over an underlying set $A$ of alternatives, a \emph{Kemeny ranking} is a linear order which minimizes the sum of distances to the $n$ linear orders, where the distance is defined via swap distance, thus equivalent to minimizing the sum of absolute distances.
\qed\end{proof}

% As for existence and uniqueness, note that a Kemeny ranking always exists but is not guaranteed to be unique; as a simple example, consider an election with only two voters $v_1$, $v_2$, and observe that both the ideal point of $v_1$ and that of $v_2$ are Kemeny rankings, thus minimizes $L_1$.
% (As another example, with odd number of voters, uniqueness is not guaranteed; e.g., consider votes $v_1 : a \pref b \pref c$, $v_2 : b \pref c \pref a$, and $v_3 : c \pref a \pref b$.)
  
As for $L_\infty$, it is equivalent to the NP-hard \textsc{Adjacent Element Swap Centre Permutation} problem~\cite[Corollary 1]{popov}. Thus $L_\infty$ is NP-hard here as well. This fact also follows from Biedl et al.~\cite{biedlhardness}.

\begin{corollary}
  For this setting, $L_\infty$ aggregation is NP-hard.
\end{corollary}

The following result shows that, for this setting, $L_p$ is NP-hard for any $p \geq 1$.

\begin{theorem}
  For this setting, $L_p$ aggregation is NP-hard for any $1 < p < \infty$.
\end{theorem}

\begin{proof}
We reduce the problem of finding the aggregation point for $L_1$ to the problem of finding the aggregation point for $L_p$, for $1 < p < \infty$. The proof then follows as $L_1$ aggregation is NP-hard (Observation~\ref{observation:kemenynphard}).

For simplicity of presentation, we describe the proof for the case of $3$ voters; but the proof generalizes in a straightforward way to more voters. Let $A$ be a set of alternative, and let $V = \{a, b, c\}$ be the voters; i.e., each of $a$, $b$, and $c$, is a permutation over $A$. Let $x^*$ be the $L_1$ aggregation point of $V$. We reduce the problem of finding $x^*$ to the problem of finding the $L_p$ aggregation point of the election we describe next.

The reduced election has a set $A \cup D \cup D'$ of alternatives (where the sets $D$ and $D'$ are sets of dummy alternatives of sufficient size; details below). Let $d$ be an arbitrary (but fixed) permutation over $D$ and $d'$ an arbitrary (but fixed) permutation over $D'$.
The reduced instance has a set $V' = \{A, B, C\}$ of voters, such that:
\begin{align*}
    A = a \pref d \pref b \pref d' \pref c \\
    B = b \pref d \pref c \pref d' \pref a \\
    C = c \pref d \pref a \pref d' \pref b
\end{align*}

Recall that $x^*$ is the $L_1$ aggregation point of $V$.
Consider the permutation
\begin{align*}
    X^* = x^* \pref d \pref x^* \pref d' \pref x^*\ .
\end{align*}
Below we show that $X^*$ minimizes the $L_p$ distance, and that only strings of this form (i.e. more generally, using any three optimal $L_1$ aggregation points of $V$ instead of three copies of $x^*$). To show this, consider an arbitrary permutation $X$ (in the metric space of the reduced election).

First, observe that, if $D$ and $D'$ are large enough, then if $X$ is not of the form $X = y_1 \pref d \pref y_2 \pref d' \pref y_3$ (where $y_i$ is a permutation over $A$), then it would incur a higher distance than $X^*$, due to the dummy alternatives (note that the dummy alternatives are disjoint from the alternatives in $A$).

So, assume that $X$ is of the form
$$X = y_1 \pref d \pref y_2 \pref d' \pref y_3\ .$$
Furthermore, we assume that $X$ does not only use $L_1$ aggregation points of $V$, i.e. that $L_1(y_1, V)+L_1(y_2, V)+L_1(y_3, V)>3 \cdot L_1(x^*, V)$.

Next we compute the $L_p$ distance between $X$ and the electorate $V'$; denote this as $L_p(X, V')$:
\begin{align*}
    L_p(X, V') &= d(X, A)^p + d(X, B)^p + d(X, C)^p \\
               &= (d(y_1, a) + d(y_2, b) + d(y_3, c))^p \\
               &+ (d(y_1, b) + d(y_2, c) + d(y_3, a))^p \\
               &+ (d(y_1, c) + d(y_2, a) + d(y_3, b))^p \\
               &= u^p + v^p + w^p\ ,
\end{align*}
where the first equality is by definition;
the second equality follows whenever $D$ and $D'$ are large enough;
and the third is simply by defining $u$, $v$, and $w$ (as notation).

Now we have the following:
\begin{align*}
    u + v + w &= d(y_1, a) + d(y_2, b) + d(y_3, c) \\
              &+ d(y_1, b) + d(y_2, c) + d(y_3, a) \\
              &+ d(y_1, c) + d(y_2, a) + d(y_3, b) \\
              &= d(y_1, a) + d(y_1, b) + d(y_1, c) \\
              &+ d(y_2, a) + d(y_2, b) + d(y_2, c) \\
              &+ d(y_3, a) + d(y_3, b) + d(y_3, c) \\
              &= L_1(y_1, V) + L_1(y_2, V) + L_1(y_3, V) \\
              &> 3 \cdot L_1(x^*, V)\ ,
\end{align*}
where the first equality is by definition of $u$, $v$, and $w$;
the second equality is by rearranging terms;
the third equality is by definition of $V$ and of $L_1$;
and the (last) inequality is from the fact that one of $y_1,y_2,y_3$ is not optimal.

Next we mention a folkore algebraic inequality on generalized means (see, e.g.,~\cite{bullen2013handbook}), stating that for positive integers $n_1, \ldots, n_k$ and $p>1$, 
$$ \frac 1k {\sum_{i \in [k]} n_i} \leq  \left( \frac 1k {\sum_{i \in [k]} n_i^p} \right)^{\frac 1p}.$$
In other words, for any $m$, we have that
$$
   \sum_{i \in [k]} n_i \geq k \cdot m \ \Rightarrow \   \sum_{i \in [k]} n_i^p \geq k \cdot m^p
$$

Using this algebraic inequality,
we have that,
since $u + v + w \geq 3 \cdot L_1(x^*, V)$,
then $L_p(X, V') = u^p + v^p + w^p  >  3 \cdot L_1(x^*, V)^p$.
Now, note that:
\begin{align*}
    3 \cdot L_1(x^*, V)^p &= 3 \cdot (d(x^*,a)+d(x^*,b)+d(x^*,c))^p \\
                          &= L_p(X^*, V')\ .
\end{align*}

We conclude that $X^*$ is the $L_p$ aggregation point.
Thus, to find the $L_p$ aggregation point, one cannot avoid finding the $L_1$ aggregation point.
\qed\end{proof}

We conjecture that also Condorcet aggregation is NP-hard for this setting, however proving it seems to be quite tricky.

\begin{conjecture}
  Condorcet aggregation for this setting is NP-hard.
\end{conjecture}

The next example demonstrates that $L_p$ aggregation is not monotone for any $p > 1$.

\begin{proposition}
  $L_\infty$ is not monotone.
\end{proposition}

\begin{proof}
The proof follows from the one-dimensional case, in particular, from Proposition~\ref{proposition:two} as we can devise instances of social welfare functions that would correspond to one-dimensional lines.

Consider, for example, two ideal points, $v_1 = a \pref b \pref c \pref d \pref e$ and $v_2 = e \pref a \pref b \pref c \pref d$. Then, observe that the ranking $w = a \pref b \pref e \pref c \pref d$ satisfies $w\in L_\infty(V)$; however, changing $v_2$ to vote as $w$ would result in a new aggregation point $a \pref b \pref c \pref e \pref d$. 
Notice that the instance described above indeed simulates a one dimensional line (containing five points, where $v_1$ corresponds to the right-most point, while $v_2$ corresponds to the left-most point).
\qed\end{proof}

Next we consider majoritarity.

\begin{proposition}
  Condorcet aggregation and $L_1$ aggregation are majoritarian for the application to social welfare functions, while $L_p$ aggregation for $p > 1$ are not.
\end{proposition}

\begin{proof}
For Condorcet aggregation, majoritarity always holds.
For $L_1$ aggregation, 
consider linear order corresponding to the voter majority's choice and, intuitively, observe that deviating from this linear order would increase the sum of absolute errors.
More formally, consider a situation with voter majority on one ideal point $x$ and consider some other ideal element $y$, of distance $d(x, y)$ from $x$. Then, each voter voting on $x$ would cause an increase to the sum of distances for $y$, which would be at least the sum of distances of all voters from $x$; thus, $x$ would minimize $L_1$.

For $L_p$ aggregation for $p > 1$, consider the set $\{a, b, c\}$ of underlying alternatives and three voters with linear orders $a \pref b \pref c$, $a \pref b \pref c$, and $c \pref a \pref b$.
Then, we have that the sum of $L_p$ distances from $a \pref b \pref c$ (which is the majority choice) is $2^p$, which is at least $4$ for $p \geq 2$, while the sum of $L_p$ distances from $a \pref c \pref b$ is $1^p + 1^p + 1^p = 3$, which is smaller.
\qed\end{proof}

\section{Committee Elections}\label{section:vnw}

Here we consider selecting a committee from a given set of candidates.

\mypara{Formal model}
Let $A$ be the underlying set of alternatives and notice that the set of possible committees is $2^A$ (that is, the set of all possible subsets of $A$). Thus, we consider the metric space $(X, d)$ where $X \subseteq 2^A$; specifically, each $x \in X$ is a subset of $A$, $x \subseteq A$.
As two specific cases, note that $X = 2^A$ corresponds to the setting where the size of the committee is not given a priori but is decided by the aggregation method itself. This scenario was considered by Kilgour~\cite{kil:j:variable-size-committee}, Duddy et al.~\cite{Duddy2016}, and Faliszewski et al.~\cite{vnw}.
However, denoting by $2^A_k$ the set of all $k$-size subsets over $A$, note that $X = 2^A_k$ corresponds to the setting of $k$-committee selection, in which $k$ committee members shall be elected. This setting is studied quite extensively~\cite{mwchapter}.
As we assume no structure or dependencies between the alternatives, we do not impose any particular structure on $X$ and define $d$ to be the \emph{symmetric distance}:
  For each $x, y \in X$, $d(x, y) = |x \triangle y|$ (indeed, $d(x, y)$ is the Hamming distance between the corresponding binary vectors of $x$ and $y$). 

First, recall the definition of the median element~\cite{reshef}:
  Consider an election $E$ based on $n$ ideal elements $V = \{v_1, \ldots, v_n\}$, $v_i \in X$. Then, the \emph{median element} is the element $x \in X$ which contains exactly those alternatives $a \in A$ for which $|\{v \in V : a \in v\}| \ge n / 2$.
That is, the median element is the unique element containing exactly those alternatives with weak voter majority support.

\begin{proposition}
  For the application to committee elections, if $X = 2^A$ then $L_1(V)$ chooses the median element, while if $X = 2^A_k$, then $L_1$ aggregation is approval voting.
\end{proposition}

\begin{proof}
For $X = 2^A$, for a set of voters $V$, the aggregated point $x = L_1(V)$ minimizes the expression $\sum_{v_i \in V} |x \triangle v_i|$, which equals to $\sum_{v_i \in V} \sum_{a \in A} \mathbbm{1}_{a \in x \triangle v_i}$, which equals to $\sum_{a \in A} \sum_{v_i \in V} \mathbbm{1}_{a \in x \triangle v_i}$.
Thus, any element $y$ different from the median element $x$ would give a higher sum, as for each $a \in x \triangle y$, $\sum_{v_i \in V} \mathbbm{1}_{a \in y \triangle v_i} \geq \sum_{v_i \in V} \mathbbm{1}_{a \in x \triangle v_i}$ holds. 
The case where $X = 2^A_k$ follows similarly.
\qed\end{proof}

\begin{remark}
The MW rule, described recently by Faliszewski et al.~\cite{vnw}, which elects a committee containing all candidates with strong majority approval, is the $L_1$ aggregation where $X = 2^A$.
Notice that $L_1(V)$ is not unique for even number of voters (e.g., consider $V = \{v_1, v_2\}$ with $v_1 : \{a\}$ and $v_2 = \{b\}$; then, both $\{a\}$ and $\{b\}$ are winners). For odd number of voters $L_1(V)$ is unique.
\end{remark}

For $X = 2^A$, finding $L_\infty(V)$ is equivalent to the NP-hard \emph{Closest String} problem~\cite{frances1997covering}. We also mention that it is equivalent to the variable number of winners counterpart of minimax approval voting~\cite{brams2007approval}.

\begin{corollary}
  For the application to committee elections with variable number of winners, computing $L_\infty(V)$ is NP-hard.
\end{corollary}

Chen et al.~\cite{chen2018computing}, using a slightly different jargon, prove that $L_p$ aggregation in NP-hard for all values of $p > 1$.

\begin{corollary}
  For the application to committee elections with variable number of winners, $L_p$ aggregation is NP-hard for any $p > 1$.
\end{corollary}

While $L_p$ aggregation always exist for this setting, the next example demonstrates that Condorcet winners do not always exist for this application.

\begin{example}
Consider alternatives $A = \{a, b, c\}$ and voters $V = \{v_1, v_2, v_3, v_4, v_5\}$ where
  $v_1 = v_2 = \emptyset$, $v_3 = \{a, b\}$, $v_4 = \{a, c\}$, and $v_5 = \{b, c\}$.
Then, $\emptyset$ beats all other points in the metric space, except $\{a, b, c\}$, but $\{a, b, c\}$ is beaten by any of $\{a\}$, $\{b\}$, and $\{c\}$; thus, no Condorcet winner exists here.
\end{example}

Condorcet aggregation for this application can be phrased also as a Condorcet-consistent aggregation method for a binary combinatorial domain with top-based input, where the completion principle is based on the Hamming distance (see, e.g.,~\cite{combinatorialcomsoc}). Condorcet aggregation was studied by Darmann~\cite{darmann2013hard} which demonstrated the computational intractability of a related problem~\cite[Theorem 3.7]{darmann2013hard}. 

We note that for committee election, only one committee may be a Condorcet winner: namely the one obtained by the majority voting rule, i.e. taking each alternative selected in a majority of preferences. Indeed, for any other committee, there is no strict majority against adding or removing an ``unpopular'' alternative. However, this is a necessary condition only, and it remains to decide if such committee actually is a Condorcet winner. 

\begin{theorem}
  For the application to committee elections with variable number of winners, Condorcet aggregation is NP-hard (more precisely, it is coNP-hard to decide whether the committee obtained by majority voting is a Condorcet winner). 
\end{theorem}

\begin{proof}
We provide a reduction from the NP-hard {\sc Vertex Cover} problem. 
%First note that {\sc Vertex Cover} is still NP-hard if we restrict ourselves to graphs for which the optimal cover is even (indeed, an easy reduction from the general case of {\sc Vertex Cover} consists in taking two disjoint copies of the input graph: then an optimal VC for the duplicated graph must have even size).
Consider a graph $G$ with vertex set denoted $A$ (with $n:=|A|$) and edge set $E$ (with $|E|$);
%--- assuming that $G$ has no odd optimal vertex cover --- 
and an even integer $k$.  We first duplicate every edge in $E$ ($E$ becomes a multi-set), and write $m$ for the (new) size of $E$. Note that this operation is safe, in the sense that a subset of vertices is a vertex cover in the original graph if, and only if, it is  a vertex cover in the new one. We further assume that $G$ has no isolated vertices, is not a star, and that $n\geq k+4$.

Let the set of alternatives be $\mathcal{A}:=A\cup B$, where $B$ is a set of $k-1$ dummy alternatives. Build a set of $2m+1$ voters $\mathcal{V}:=E \cup F \cup \{g\}$ as follows:
\begin{itemize}
    \item $E$ is the set of edges, i.e. it is a set of $m$ voters with preference $\{u,v\}\subset A$
    \item $F$ is a set of $m$ voters, all preferring set $B$
    \item $g$ is a voter with preference $\mathcal{A}=A\cup B$    
\end{itemize}

The majority voting winner for this election (denoted $w$) is the set $B$ of alternatives. Indeed, each element of $B$ appears in the preference set of $m+1$ voters, and no element in $A$ appears in $m+1$ preference sets (otherwise such an element would be a vertex incident to all edges, i.e. $G$ would be a star).
We prove the following equivalence:
$$w \text{ is not a Condorcet winner} \Leftrightarrow G \text{ has a vertex cover of size } k.$$

We first compute the distances between $w$ and the different voters. For $e\in E$, $d(w,e)=|B|+|e|=k+1$ (since $e$ and $B$ are disjoint). 
For $f \in F$, $d(w,f)=0$. 
Finally, for voter $g$, $d(w,g)=|A|=n$. 
Now for any other committee $h\in 2^{\mathcal A}$, $h\neq w$, we have $d(h,e)=|h|-2$ if $h$ contains $e$, $d(h,e)=|h|$ if $h$ contains one endpoint of $e$, and $d(h,e)=|h|+2$ otherwise. Also, $d(h,f)>d(w,f)$ for all $f\in F$, and $d(h,g)<d(w,g)\Leftrightarrow |h|\geq k$.

$\Leftarrow$. Let $h\subset A$ be a vertex cover of $G$ of size $k$ (then $h\subset \mathcal A$ is also a candidate committee). For any voter $e\in E$, since $h\cap e\neq \emptyset$, we have $d(h,e)\leq |h| = k <d(w,e)$; and for voter $g$, $d(h,g)<d(w,g)$. So all $m+1$ voters in $E\cap\{g\}$ strictly prefer $h$ over $w$: $w$ is not a Condorcet winner.

$\Rightarrow$. %Aiming at a contradiction, 
Assume that $w$ is not a Condorcet winner. %, and that $G$ has no vertex cover of size $k$. 
Let $h\subseteq A\cup B$, $h\neq w$ be a committee beating $w$, i.e. a (not necessarily strict) majority of voters prefer $h$ over $w$. Since voters in $F$ strictly prefer $w$ over $h$, then all voters in $E\cup\{g\}$ must strictly prefer $h$ over $w$ (except at most one voter, in case of a tie, which is at the same distance from both). Thus, $|h|>k-1$, and $d(h,e)<k+1$ for all $e\in E$ (except for at most one strict inequality which may be an equality). In particular, using $d(h,e)\geq |h|-2$, the latter implies $|h|< k+3$.

First if $|h|=k-1$ (the unique equality is for voter $g$), then $d(h,e)<k+1$ implies that $h$ intersects every edge $e$: $h$ is a size-$(k-1)$ vertex cover.

Now if $|h|=k$, and $e\cap h\neq \emptyset$ for all (or all but one) edge $e$. Then $h$ covers at least $m-1$ edges, which is sufficient to be a vertex cover (since every edge has been duplicated).

Finally if $|h|\geq k+1$, then $d(h,e)=|h|-2$ for all but one edge, i.e. $e\subset h$. Thus,  $h$ contains all vertices (except at most one single pending vertex) and $|h|\geq n-1$. Using $|h|< k+3$, we get $n<k+4$, a contradiction.
%
%In particular for $g$, it means that $|h|\geq k$. For each $e$ in $E$, $d(h,e)\leq k$. If $|h|>k$, then $e\subset h$ for every $e$ and $|h|\leq k+2$ (i.e. $n\leq k+2$ if $G$ is connected: a contradiction). Hence $|h|=k$ and $e\cap h\neq \emptyset$ for every edge $e$. Thus, $h\cap A$ is a vertex cover of size at most $k$.
%
%Now $h\cap B$ is not a size-$k$ vertex cover, so either it has size $k+1$, or it does not cover some vertex $e\in E$. In the first case, $d(h,e)\geq k +1$ for any $e\not\subset h$, and in the second case $d(h,e)\geq k+2$ for the uncovered edge $e$. In both cases, some voter in $E$ does not strictly prefer $h$ over $w$: a contradiction.
\qed
\end{proof}

Next we consider majoritarity.

\begin{observation}
  For the application to committee elections with variable number of winners, Condorcet and $L_1$ aggregation are both majoritarian,
  while $L_p$ aggregation for $p > 1$ are not majoritarian.
\end{observation}

\begin{proof}
For Condorcet aggregation majoritarity holds for any metric space, thus, in particular, also here. For $L_1$ aggregation, the claim follows as the median element would be the choice of the voter majority, and, in case of voter majority on one ideal element, it would be the median element.

For $L_p$ aggregation with $p > 1$, choose sufficiently large odd $n$ such that $\frac{n}{n-1} < 2^{p-1}$. Now, consider the set $A = \{a, b\}$ of underlying alternatives, and a set of votes given by $V= \{ \{a\}^{(n+1)/2}, \{b\}^{(n-1)/2} \}$, where the exponent of each point corresponds to the number of votes for this point.

For this instance, we have

$$min_{x\in X} \sum_{i \in [n]} d(v_i, x)^p \leq \sum_{i \in [n]} d(v_i, \{a,b\})^p = n\ .$$

In contrast, 

$$\sum_{i \in [n]} d(v_i, \{a\})^p = \frac{n-1}{2}\cdot 2^p = (n-1)2^{p-1} > n\ .$$

It follows that, even though the majority choice is $\{a\}$, the $L_p$ aggregation of this instance must be some $x\neq a$ in $X$.
\qed\end{proof}

The following example shows that monotonicity is not satisfied for $p > 1$.

\begin{example}
Consider an underlying set $A = \{a, b, c, d\}$ of alternatives and ideal points $V = \{v_1, v_2\}$ with $v_1 = \emptyset$ and $v_2 = A = \{a, b, c, d\}$.
Then, the point $\{a, b\}$ is an aggregation point for $L_p$ aggregation, for $p > 1$, however if $v_2$ moves to $\{a, b\}$ then the new aggregation point would be $\{a\}$.
\end{example}

%\begin{corollary}
 % Monotonicity is not satisfied by $L_p$ aggregation, for any $p > 1$, for the application to committee elections with variable number of winners.
%\end{corollary}

Condorcet aggregation does satisfy monotonicity, as it does for all applications we consider in this paper. $L_1$ aggregation is monotone here as the median element would stay when moving a voter ideal point to it.

\section{Participatory Legislation}\label{section:legislation}

We consider participatory legislation (see, e.g.,~\cite{alsina2018birth}). Clearly, plurality elections can be applied to approve a proposed legislation, or to choose among a given set of proposed legislation.  Even if such a proposal is made by an elected legislation committee, such a process falls short of the egalitarian vision of the 1789 Declaration of the Rights of Man and Citizen~\cite[Article VI]{french}, which states that all citizens have the right of contributing to the formation of the law personally, not only via their representatives. Hence, we focus here on aggregating text drafts, e.g., to jointly write a constitution. In fact, our main motivation for this work comes from the apparent lack of literature on a participatory approach to this key social choice setting in the computational social choice community.

\mypara{Formal model}
Here we have a certain alphabet $\Sigma$. The possible outcomes, as well as the possible votes, are all strings over $\Sigma$; thus, we consider a metric space $(X, d)$ where $X = \Sigma^*$, such that each $x \in X$ is a string over the alphabet $\Sigma$.

Various possibilities exist for defining the metric $d$. The first thing that comes to mind is to consider the Levenstein edit distance:
  For two ideal element $x, y \in X$, the Levenstein distance $d(x, y)$ is the minimum number of edit operations of the form $insert$ (which adds a character at a certain position), $delete$ (which deletes a character at a certain position), and $swap$ (which swaps two consecutive characters) that shall be performed to convert $x$ into $y$.
Assume that strings are documents consisting of a sequences of sentences, namely that characters are sentences. We feel that in this setting, this metric is not appropriate as it gives the same importance to the identity of the sentences as to their order.
However, when writing, e.g., a constitution, the contents of the sentences in it are more important than their order.
For this reason,
we consider a different distance metric, namely a weighted-Levenstein distance $d(x, y)$, which equals the minimum weight of operations that transform $x$ to $y$, with the following weighted operations:
  an $insert$ operation costs $1$,
  a $delete$ operation costs $1$,
  and
  a $swap$ operation costs $1/\ell^2$,
  where $\ell$ is the length of the longest ideal point proposed by any voter.
The situation now boils down to a \emph{two-phase} election:
  First, one shall find the set of sentences to be included in the constitution;
  second, one shall find the best ordering of these sentences.
We consider these two phases in turn.

\mypara{Phase 1: Selecting the set of sentences}
Here, only considering the set of sentences in each ideal point, the task is to aggregate those $n$ sets of sentences into one set of sentences. We assume that the resulting aggregated has at most one copy of each sentence. Thus, this setting is formally equivalent to multiwinner elections with variable number of winners, studied in Section~\ref{section:vnw}.

\mypara{Phase 2: Selecting the order of the selected sentences}
Following phase 1, we shall order the elected sentences. This setting is formally equivalent to social welfare functions, studied in Section~\ref{section:swf}.

\mypara{Combining the two phases}
Naturally, we get the combined hardness of the two phases: Existence is guaranteed only for $L_p$ aggregation; uniqueness is not guaranteed for any aggregation method; Condorcet and $L_1$ aggregation are majoritarian, while $L_p$ aggregation for $p > 1$ are not majoritarian; and computational hardness follows.

\begin{remark}
For $L_1$ aggregation, one can use the polynomial time algorithm for $L_1$ aggregation for multiwinner elections with variable number of winners discussed in Section~\ref{section:vnw}, followed by solving Phase 2 heuristically, e.g., using one of the methods for Kemeny ranking studied by Ali and Meil{\u{a}}~\cite{ali2012experiments}).
\end{remark}

\section{Discussion}\label{section:outlook}

We proposed a general metric-based model and applied it to a broad range of social choice settings, demonstrating its applicability as providing a unifying elicitation and aggregation framework for the social choice situations needed for an e-democracy.
Below we discuss some avenues for future research.

\subsection{Further Settings}
Further social choice settings shall be realized in our general model. In particular, further distance measures, such as, e.g., those described by Schiavinotoo and St{\"u}tzle~\cite{schiavinotto2007review} can be applied.
Related, our aggregation methods for participatory legislation only take into account syntactic distance, but not semantic distance. An intriguing avenue for future research would be to apply NLP methods which take into account the semantics of the aggregated text drafts.

\subsection{Extensions}
A major advantage of utilizing a unified framework, capturing many social choice settings at once, is that extensions and generalizations can be designed in a unified way, making them applicable to all such settings at once.
One natural extension would be to incorporate proxy voting~\cite{miller1969program} and liquid democracy~\cite{liquidfeedback,brill2018pairwise} to the model.
Another natural extension would be to incorporate sybil resilience~\cite{SRSC} to the model, making it applicable in situations in which not all agents participating in the process can be assumed to be genuine.

\subsection{Voting and Deliberation}
A major caveat of our model is that we infer voter preferences over the metric elements based only on their ideal points, while it might be the case that, e.g., voters with identical ideal points have different preferences over other elements of the metric space.
This is especially problematic, e.g., when dependencies exist between committee members in multiwinner elections; where interdependent items are to be funded in participatory budgeting; or where different sentence ordering alter the meaning of the resulting text.

One remedy for this might be to incorporate Reality (the status quo) in a process that interleaves voting with deliberation; for instance, similarly in spirit to Reality-aware action plans~\cite{realsoc} and to other works~\cite{iterativeja}, we envision an iterative process where the deliberation phase of each iteration is based on the aggregated point identifying based on the voting phase of the last iteration, following which participants may change their votes; the result of such an iterative process could be an aggregated point that has majority support over the status quo, or else a recognition that the status quo is presently preferred over any aggregated alternative.

\section*{Acknowledgements}

We thank the generous support of the Braginsky Center for the Interface between Science and the Humanities.
Nimrod Talmon was supported by the Israel Science Foundation (ISF; Grant No. 630/19).

\bibliographystyle{splncs04}
\bibliography{bib}

\end{document}